\documentclass{ws-mpla_blank} 
\usepackage{graphicx}         

\newcommand{\fracnew}[2]
           {\protect\frac{{#1}_{\protect\vphantom{!_A}}}
           {{#2}^{\protect\vphantom{A}}}}

\begin{document}

\title{\vspace*{-1.25cm}BLACK-HOLE SOLUTION WITHOUT CURVATURE SINGULARITY}

\author{F.R. Klinkhamer}
\address{Institute for Theoretical Physics,
Karlsruhe Institute of Technology (KIT),\\
76128 Karlsruhe, Germany\\
frans.klinkhamer@kit.edu}

\maketitle
\begin{abstract}
An exact solution of the vacuum Einstein field equations over
a nonsimply-connected manifold is presented.
This solution is spherically symmetric and has no curvature singularity.
It can be considered to be a regularization of the Schwarzschild solution
over a simply-connected manifold, which has a curvature singularity
at the center.
Spherically symmetric collapse of matter in $\mathbb{R}^4$ may
result in this nonsingular black-hole solution,
if quantum-gravity effects allow for topology change near the center.%
\vspace*{1.25\baselineskip}\newline
Journal: \emph{Mod. Phys. Lett. A} \textbf{28}, 1350136 (2013)
\vspace*{.25\baselineskip}\newline
Preprint:  arXiv:1304.2305
\vspace*{.25\baselineskip}\newline
\vspace*{.25\baselineskip}\newline
\end{abstract}

\maketitle

\section{Introduction}
\label{sec:Introduction}

The vacuum Einstein field equations over $\mathcal{M}_4=\mathbb{R}^4$
have a spherically symmetric solution,
the Schwarzschild black-hole solution\cite{Schwarzschild1916}
with its maximal extension.\cite{Kruskal1960,Szekeres1960,MTW1974}
Recently, a related solution has been found for
the vacuum Einstein field equations over
$\widetilde{\mathcal{M}}_4$, a particular nonsimply-connected
manifold.\cite{KlinkhamerRahmede2013}

The goal of the present article is to
discuss the connection between these two vacuum solutions.
The comparison of these two solutions will, in fact, suggest
a black-hole-type solution, where the curvature singularity
is eliminated by a spacetime defect, i.e., a ``hole'' in spacetime.

\section{Singular black-hole solution}
\label{sec:Singular-black-hole-solution}

The extended Schwarzschild--Kruskal--Szekeres
solution\cite{Schwarzschild1916,Kruskal1960,Szekeres1960}
has a metric given by the following line element ($G_N=c=1$):
\begin{subequations}\label{eq:SKS-solution}
\begin{equation}\label{eq:SKS-solution-ds2}
ds^2\,\Big|^{(r\, >\, 0)}_\text{SKS}=
32\, M^3\; \frac{\exp\big[-r/(2M)\big]}{r}\;
\Big( -dv^2 +du^2\Big)
+ r^2\, \Big( d\theta^2 +\sin^2\theta\; d\phi^2 \Big)\,,
\end{equation}
with $\theta\in [0,\,\pi]$, $\phi\in [0,\,2\pi)$,
and $r$ implicitly given in terms of the
coordinates $u\in \mathbb{R}$ and $v\in \mathbb{R}$ by the relation
\begin{equation}\label{eq:SKS-solution-r-implicit}
\left(\frac{r}{2M} -1 \right)\;
\exp\left[\frac{r}{2M}\right]
\;\Big|_\text{SKS}
= u^2 -v^2 > -1\,,
\end{equation}
for $r>0$. The explicit expression for $r$ from \eqref{eq:SKS-solution-r-implicit}
is
\begin{equation}\label{eq:SKS-solution-r-explicit}
r\;\Big|_\text{SKS} =
2M \left( 1 + W_{0}\left[\frac{u^2 -v^2}{e}\right]\right)\,,
\end{equation}
\end{subequations}
in terms of the principal branch of the
Lambert $W$--function,  
$W_{0}[z]$, which gives the principal solution
for $w$ in \mbox{$z=w \,\exp[w]$.}
Recall that $W_{0}[x]$ is real for $x \geq -1/e$.

The solution \eqref{eq:SKS-solution} has $M>0$ as a free parameter
and the corresponding topology is
\begin{equation}
\mathcal{M}_\text{SKS} =\mathbb{R}^2 \times S^2\,.
\end{equation}
The solution approaches a physical singularity for $r\to 0$,
as shown by the divergence of the Kretschmann scalar
(defined in terms of the Riemann curvature tensor),
\begin{eqnarray}\label{eq:Kscalar-SKS-solution}
K\;\Big|_\text{SKS}
&\equiv&
R_{\mu\nu\rho\sigma}\,R^{\mu\nu\rho\sigma}\;\Big|_\text{SKS}
=48\;\frac{M^2}{r^6}\,.
\end{eqnarray}
Further details can be found in, e.g., Ref.~\refcite{MTW1974}.

\section{Nonsingular black-hole solution}
\label{sec:Nonsingular-black-hole-solution}

Consider, now, the vacuum Einstein field equations over
\begin{subequations}\label{eq:M4-product-M3}
\begin{eqnarray}\label{eq:M4-product}
\widetilde{\mathcal{M}}_4 &=& \mathbb{R} \times \widetilde{\mathcal{M}}_3 \,,
\end{eqnarray}
where the 3-space $\widetilde{\mathcal{M}}_3$ is  a noncompact,
orientable, nonsimply-connected manifold without boundary.
Specifically, there is the following homeomorphism:
\begin{eqnarray}\label{eq:M3}
\widetilde{\mathcal{M}}_3
&\simeq&
\mathbb{R}P^3 - \{\text{point}\}\,,
\end{eqnarray}
\end{subequations}
with $\mathbb{R}P^3$ the 3-dimensional real projective plane,
which is topologically equivalent to a 3-sphere with antipodal
points identified. Here, and in the following, $\widetilde{\mathcal{M}}$
with tilde indicates a nonsimply-connected manifold
[having a nontrivial first homotopy group,
$\pi_1(\widetilde{\mathcal{M}})\ne 0$], whereas $\mathcal{M}$ without
tilde stands for a simply-connected manifold
[having $\pi_1(\mathcal{M})=0$].

The explicit construction of $\widetilde{\mathcal{M}}_3$
has been given in Ref.~\refcite{KlinkhamerRahmede2013}:
start from $\mathbb{R}^3$ with the standard
Cartesian coordinates and the Euclidean metric and,
then, perform surgery on $\mathbb{R}^3$
by removing the interior of a ball $B_b$ with radius $b$
and identifying antipodal points on the boundary
of the ball, $\partial B_b=S_b^2$ (2-sphere with radius $b$).
In Ref.~\refcite{KlinkhamerRahmede2013},
also appropriate coordinate charts are reviewed. There is,  for example,
one chart (labeled $n=2$) with `radial' coordinate
$Y_{2}\in (-\infty,\,+\infty)$ instead of the standard radial coordinate
$r\in [b,\,\infty)$
and angular coordinates $(X_{2},\, Z_{2})$ having restricted ranges.
(The fundamentally different nature of the
coordinate $Y_2$  compared to that of the coordinate $r$
will play an important role later on.)
For the moment, the focus will be on these chart-2 coordinates
and the subscript `$2$'  will simply be omitted.

The new exact vacuum solution over $\widetilde{\mathcal{M}}_4$
involves two parameters, $b$ and $M$, which, for the present purpose,
are taken to be related as follows:
\begin{equation}\label{eq:b-less-than-2M-condition}
0 < b < 2 M\,.
\end{equation}
The defect solution for $2 M <b$
has been given by Eqs.~(4.1), (5.1b), and (5.1c)
in Ref.~\refcite{KlinkhamerRahmede2013} for chart-2 coordinates
and inspection shows it to be precisely of the form
of the standard exterior-region Schwarzschild solution
if $Y^2+b^2$ and $r^2$ are identified.
This observation agrees with Birkhoff's theorem\cite{MTW1974}
and, moreover,  allows us to use some of the
techniques of the Kruskal--Szekeres procedure
in our search for a solution with parameters
\eqref{eq:b-less-than-2M-condition}.

It turns out that the Kruskal--Szekeres procedure is
useful to eliminate the apparent singularities
at the Schwarzschild horizon but is not suited to deal
with the spacetime defect.  For this reason, we will present our
black-hole-type solution without curvature singularity
in terms of two sets of coordinates, one
set of coordinates appropriate to the spacetime defect
and another set of coordinates further out.

Start, then, with an
\textit{Ansatz} in terms of Kruskal--Szekeres-type coordinates,
valid for $\zeta > b$ and having a metric
given by the following line element:
\begin{subequations}\label{eq:new-BH-solution-UVcoord}
\begin{equation}\label{eq:new-BH-solution-UVcoord-ds2}
ds^2\,\Big|^{(\zeta \, > \, b)}=
32\, M^3\; \frac{\exp\big[-\zeta/(2M)\big]}{\zeta}\;
\Big( -d V^2 +d U^2\Big)
+ \zeta^2\, \Big( dZ^2 +\sin^2Z\; dX^2 \Big)\,,
\end{equation}
with $\zeta$ given in terms of
$U\in \mathbb{R}$ and $V\in \mathbb{R}$ by
\begin{equation}\label{eq:new-BH-solution-UVcoord-zeta}
\zeta =
2M \left( 1 + W_{0}\left[\frac{U^2 -V^2}{e}\right]\right)
=\sqrt{b^2+Y^2}\,,
\end{equation}
where the last equality follows from the relations
\begin{eqnarray}\label{eq:new-BH-solution-coordinate-U}
U  &=&
\left\{
\begin{array}{c}
\left(\fracnew{\sqrt{\textstyle{Y^2+b^2}}}{\textstyle{2M}}-1\right)^{1/2}\;
\exp\left[\fracnew{\sqrt{\textstyle{Y^2+b^2}}}{\textstyle{4M}}\right]\,
\cosh\left[\fracnew{\textstyle{T}}{\textstyle{4M}}\right] \,,
 \\[5mm]
\left(1-\fracnew{\sqrt{\textstyle{Y^2+b^2}}}{\textstyle{2M}}\right)^{1/2}\;
\exp\left[\fracnew{\sqrt{\textstyle{Y^2+b^2}}}{\textstyle{4M}}\right]\,
\sinh\left[\fracnew{\textstyle{T}}{\textstyle{4M}}\right]  \,,
\end{array}
\right.
\\[4mm]
\label{eq:new-BH-solution-coordinate-V}
V
&=&
\left\{
\begin{array}{c}
\left(\fracnew{\sqrt{\textstyle{Y^2+b^2}}}{\textstyle{2M}}-1\right)^{1/2}\;
\exp\left[\fracnew{\sqrt{\textstyle{Y^2+b^2}}}{\textstyle{4M}}\right]\,
\sinh\left[\fracnew{\textstyle{T}}{\textstyle{4M}}\right]\,,
 \\[5mm]
\left(1-\fracnew{\sqrt{\textstyle{Y^2+b^2}}}{\textstyle{2M}}\right)^{1/2}\;
\exp\left[\fracnew{\sqrt{\textstyle{Y^2+b^2}}}{\textstyle{4M}}\right]\,
\cosh\left[\fracnew{\textstyle{T}}{\textstyle{4M}}\right]\,,
\end{array}
\right.
\end{eqnarray}
\end{subequations}
taking the top/bottom entries in \eqref{eq:new-BH-solution-coordinate-U}
and \eqref{eq:new-BH-solution-coordinate-V} for the exterior/interior regions
(labeled I/II in Sec.~31.5 of Ref.~\refcite{MTW1974}).

Next, write
the \textit{Ansatz} \eqref{eq:new-BH-solution-UVcoord-ds2} in
terms of the chart-2 coordinates $\{Y,\,T\}\in (-\infty,\,+\infty)$
by using \eqref{eq:new-BH-solution-UVcoord-zeta}
and the bottom-row expressions on the right-hand sides
of \eqref{eq:new-BH-solution-coordinate-U}
and \eqref{eq:new-BH-solution-coordinate-V}.
In this way, it is possible to obtain the
interior metric with the defect at $Y=0$ \emph{included},
\begin{subequations}\label{eq:new-BH-solution-TYcoord-ds2}
\begin{eqnarray}\label{eq:new-BH-solution-TYcoord-ds2-interior}
ds^2\,\Big|_\text{chart-2}^{(0 \, \leq \, Y^2 \, < \, 4M^2-b^2)}
&=&
+ \left(\frac{2 M}{\sqrt{b^2+Y^2}}-1\right)\;dT^2
\nonumber\\[1mm]
&&
-  \left(\frac{2 M}{\sqrt{b^2+Y^2}}-1\right)^{-1} \;
\frac{Y^2}{b^2+Y^2}\;\;dY^2
\nonumber\\[1mm]
&&
+ (b^2+Y^2)\, \Big( dZ^2 +\sin^2Z\; dX^2 \Big)\,.
\end{eqnarray}
Similarly, one obtains from the identical \textit{Ansatz}
\eqref{eq:new-BH-solution-UVcoord-ds2}, but now
using the top-row expressions on the right-hand sides
of \eqref{eq:new-BH-solution-coordinate-U}
and \eqref{eq:new-BH-solution-coordinate-V}, the exterior metric,
\begin{eqnarray}\label{eq:new-BH-solution-TYcoord-ds2-exterior}
ds^2\,\Big|_\text{chart-2}^{(Y^2 \, >\,  4M^2 - b^2)}
&=&
- \left(1-\frac{2 M}{\sqrt{b^2+Y^2}} \right)\;dT^2
\nonumber\\[1mm]
&&
+  \left(1-\frac{2 M}{\sqrt{b^2+Y^2}} \right)^{-1} \;
\frac{Y^2}{b^2+Y^2}\;dY^2
\nonumber\\[1mm]
&&
+ (b^2+Y^2)\, \Big( dZ^2 +\sin^2Z\; dX^2 \Big)\,,
\end{eqnarray}
which takes the same form as  \eqref{eq:new-BH-solution-TYcoord-ds2-interior}.
For the $\zeta = 2M$ boundary between the exterior and interior spacetime regions
mentioned below \eqref{eq:new-BH-solution-coordinate-V},
the \textit{Ansatz} \eqref{eq:new-BH-solution-UVcoord-ds2} in terms of
$(V\,,U)$ coordinates can be kept as it is,
\begin{eqnarray}\label{eq:new-BH-solution-TYcoord-ds2-boundary}
ds^2\,\Big|_\text{chart-2}^{(2M-\Delta L\, <\,  \zeta \, <\, 2M+\Delta L)}
&=&
32\, M^3\; \frac{\exp\big[-\zeta/(2M)\big]}{\zeta}\;
\Big( -d V^2 +d U^2\Big)
\nonumber\\[1mm]
&&
+ \zeta^2\, \Big( dZ^2 +\sin^2Z\; dX^2 \Big)\,,
\end{eqnarray}
\end{subequations}
where $\zeta$ is now given by \eqref{eq:new-BH-solution-UVcoord-zeta}
and $\Delta L$ is a small enough positive length, so that $b < 2M-\Delta L$.
The coordinate transformations for the overlap regions
between \eqref{eq:new-BH-solution-TYcoord-ds2-boundary}
and  \eqref{eq:new-BH-solution-TYcoord-ds2-interior}
or \eqref{eq:new-BH-solution-TYcoord-ds2-exterior}
are given by \eqref{eq:new-BH-solution-coordinate-U}
and \eqref{eq:new-BH-solution-coordinate-V}.
Note that, strictly speaking, \eqref{eq:new-BH-solution-TYcoord-ds2-exterior}
could be omitted if the range of
\eqref{eq:new-BH-solution-TYcoord-ds2-boundary} is extended
to $\zeta \in (b,\,\infty)$, with the case $\zeta =b$ covered by
\eqref{eq:new-BH-solution-TYcoord-ds2-interior}.
But, for now, we continue with
\eqref{eq:new-BH-solution-TYcoord-ds2} as it stands.

The timelike coordinate $T$ and spacelike coordinate $Y$
of the exterior metric
\eqref{eq:new-BH-solution-TYcoord-ds2-exterior}
become, respectively, spacelike and timelike coordinates
of the interior metric
\eqref{eq:new-BH-solution-TYcoord-ds2-interior}.
This behavior is analogous
to what happens for the standard Schwarzschild solution
(cf. Fig.~31.1 in Ref.~\refcite{MTW1974}).
Note, however, that the timelike coordinate $Y$ of the
interior metric \eqref{eq:new-BH-solution-TYcoord-ds2-interior}
ranges from $-\infty$ to $+\infty$, unlike the usual radial coordinate $r$.
Moreover, this timelike coordinate $Y$ is part of a topologically
nontrivial manifold (see below), which may have important physics
implications as will be discussed in Sec.~\ref{sec:Discussion}.

The Riemann curvature
tensor $R^{\kappa}_{\;\;\lambda\mu\nu}(T,\,X,\, Y,\, Z)$
from \eqref{eq:new-BH-solution-TYcoord-ds2-interior}
is found to be even in $Y$ and finite at $Y=0$;
see the explicit expressions given by Eq.~(5.4)
in Ref.~\refcite{KlinkhamerRahmede2013}.
The Ricci tensor $R_{\mu\nu}(T,\,X,\, Y,\, Z)$
from \eqref{eq:new-BH-solution-TYcoord-ds2}
vanishes identically  and the same holds for
the Ricci scalar $R(T,\,X,\, Y,\, Z)$.
Hence, the vacuum Einstein equations are solved.

With the Riemann tensor of the vacuum
solution \eqref{eq:new-BH-solution-TYcoord-ds2},
the Kretschmann scalar is found to be given by
\begin{eqnarray}\label{eq:Kscalar-K-solution}
K  &\equiv& R_{\mu\nu\rho\sigma}\,R^{\mu\nu\rho\sigma}
=48\;\frac{M^2}{\zeta^6}\,,
\end{eqnarray}
with $\zeta^2 \equiv b^2+Y^2$ for the interior
metric \eqref{eq:new-BH-solution-TYcoord-ds2-interior}
or the exterior metric \eqref{eq:new-BH-solution-TYcoord-ds2-exterior}
and $\zeta$ given by \eqref{eq:new-BH-solution-UVcoord-zeta}
for the boundary metric \eqref{eq:new-BH-solution-TYcoord-ds2-boundary}.
For all three metrics from \eqref{eq:new-BH-solution-TYcoord-ds2},
the Kretschmann scalar remains finite because $b>0$.

As discussed in Sec.~2 of Ref.~\refcite{KlinkhamerRahmede2013}, the
proper description of $\widetilde{\mathcal{M}}_3$ requires three coordinate charts.
In the present context, there are, then, the coordinates
$(V_{n},\,U_{n},\,X_{n},\,Z_{n})$
and $(T_{n},\,X_{n},\,Y_{n},\,Z_{n})$, for $n=1,2,3$,
with
$(V_{2},\,U_{2},\,X_{2},\,Z_{2})$ $=$
$(V,\,U,\,X,\,Z)$
as used in \eqref{eq:new-BH-solution-TYcoord-ds2-boundary}
and $(T_{2},\,X_{2},\,Y_{2},\,Z_{2}) \equiv (T,\, X,\, Y,\, Z)$
as used in  \eqref{eq:new-BH-solution-TYcoord-ds2-interior} and
\eqref{eq:new-BH-solution-TYcoord-ds2-exterior}. Hence,
the interior spacetime manifold from \eqref{eq:new-BH-solution-TYcoord-ds2-interior} 
extended to all charts and
the exterior spacetime manifold from \eqref{eq:new-BH-solution-TYcoord-ds2-exterior}
extended to all charts
have the same topology, namely, $\widetilde{\mathcal{M}}_4$ from
\eqref{eq:M4-product}
in terms of $\widetilde{\mathcal{M}}_3$ from \eqref{eq:M3}.

\section{Discussion}
\label{sec:Discussion}

The solution \eqref{eq:new-BH-solution-TYcoord-ds2}
of the  vacuum Einstein field equations for the topology
\eqref{eq:M4-product-M3} is the main result of this article.
The metrics \eqref{eq:new-BH-solution-TYcoord-ds2-exterior}
and \eqref{eq:new-BH-solution-TYcoord-ds2-boundary}
are, of course, not new, as these metrics correspond to the standard
Schwarzschild--Kruskal--Szekeres metric.
New is the interior metric \eqref{eq:new-BH-solution-TYcoord-ds2-interior}
for the coordinate $Y \in (-\infty,\,\infty)$ of the
nonsimply-connected manifold $\widetilde{\mathcal{M}}_3$,
as this metric describes the spacetime defect at $\zeta = \sqrt{b^2+Y^2}=b$,
which effectively eliminates the curvature singularity at $\zeta = 0$.

Radial geodesics $Y(T)$ passing through $Y=0$
have been obtained numerically and analytically
for the metric \eqref{eq:new-BH-solution-TYcoord-ds2-interior}
with parameters $0<b<2M$.
The leading functional behavior of these radial geodesics at $Y=0$
is found to be the same as that of the radial geodesics for the flat
defect metric with parameters $M=0$ and $b>0$.

For the spacetime region inside the Schwarzschild event horizon,
$Y^2<4M^2-b^2$, the  coordinate $Y$
from the metric \eqref{eq:new-BH-solution-TYcoord-ds2-interior}
is timelike and the  coordinate $T$ spacelike, similar to
what happens for the standard Schwarzschild metric.
Now, however, this timelike coordinate $Y$
is part of the manifold $\widetilde{\mathcal{M}}_3$
from \eqref{eq:M3}, which gives rise to closed time-like curves (CTCs).
These CTCs imply all possible horrors, but, classically,
these horrors remain confined within the Schwarzschild horizon.
Whether or not CTCs in the interior region
are physically acceptable depends on the behavior of the matter fields.
Further discussion of this issue can be found in
Ref.~\refcite{Klinkhamer2013b}.
At this moment, let us simply
return to the vacuum solution \eqref{eq:new-BH-solution-TYcoord-ds2}.

Purely mathematically, the nonsingular vacuum
solution \eqref{eq:new-BH-solution-TYcoord-ds2} with parameter $b>0$
can be considered to be a ``regularization'' of part of
the singular vacuum solution \eqref{eq:SKS-solution},
with all the surprises this may entail.
It is, however, also possible that this nonsingular solution
appears in a physical context.

Start from a nearly flat spacetime
(metric approximately equal to the Minkowski metric and trivial
topology $\mathbb{R}^4$), where
a large amount of matter with total mass $M$
is arranged to collapse in a spherically symmetric way.
Within the realm of classical Einstein gravity, one expects
to end up with part of the singular solution \eqref{eq:SKS-solution};
see, for example, Fig.~32.1.b of Ref.~\refcite{MTW1974}.
But, very close to the final curvature singularity,
something else may happen due to quantum mechanics.

Considering a precursor mass $\Delta M \sim \hbar/(b\,c) \ll M$
and using typical curvature values
from \eqref{eq:Kscalar-SKS-solution} and \eqref{eq:Kscalar-K-solution},
the local spacetime integral of the action density
related to \eqref{eq:SKS-solution}
differs from that related to \eqref{eq:new-BH-solution-TYcoord-ds2}
by an amount $\lesssim \hbar$. Then, as argued by Wheeler
(cf. Secs.~34.6, 43.4, and 44.3 of Ref.~\refcite{MTW1974} and references therein),
the local topology of the manifold may change by a quantum jump
if $b$ is sufficiently close
to the Planck length $L_\text{P}\equiv(\hbar\,G_N/c^3)^{1/2}$,
resulting in a transition from a simply-connected
manifold to a nonsimply-connected manifold.
If the transition amplitude between the different topologies
is indeed nonzero for appropriate matter
content, quantum mechanics can operate a change between
the classical solution \eqref{eq:SKS-solution}
and the classical solution \eqref{eq:new-BH-solution-TYcoord-ds2},
thereby removing the curvature singularity.

\section*{Note Added}
\noindent
The flat spacetime metric \eqref{eq:new-BH-solution-TYcoord-ds2-interior}
for the case $M=0$ and $b\ne 0$
does not describe a smooth Lorentzian manifold.
There is no diffeomorphism (invertible one-to-one $C^\infty$ function)
which maps the spacetime near $Y=0$ to a part of Minkowski spacetime.
The same holds for the general case,
$M\in \mathbb{R}$ and $b\ne 0$.
This violation of the elementary-flatness condition appears to be
one of the surprises from the ``regularization,''
as mentioned in the fourth paragraph of Sec.~\ref{sec:Discussion}.

\section*{Acknowledgments}
\noindent
It is a pleasure to acknowledge discussions with
S.~Antoci, D. Grumiller,  C.~Rahmede, and H.~Sahlmann.
Further discussions with the
participants of the Karl Schwarzschild Meeting on Gravitational Physics
(Frankfurt Institute for Advanced Studies, July 2013)
have also been most helpful,
especially regarding the issues mentioned in the Note Added.
This work has been supported, in part,
by the ``Helmholtz Alliance for Astroparticle Physics (HAP),''
funded by the Initiative and Networking Fund of the Helmholtz Association.


\end{document}